\documentclass[prd,showpacs,showkeys,nofootinbib,floatfix, fleqn,preprint,12pt,tightenlines]{revtex4-2}

\usepackage{amsmath,amssymb,revsymb,graphicx,dcolumn,amsfonts}
\usepackage{marginnote}
\usepackage{mathrsfs}
\usepackage{caption}
\usepackage{subcaption}
\usepackage{cleveref}
\crefname{figure}{Fig.}{Figs.}

\newcommand{\beq}{\begin{equation}}
\newcommand{\eeq}{\end{equation}}
\newcommand{\beqa}{\begin{eqnarray}}
\newcommand{\eeqa}{\end{eqnarray}}
\newcommand{\bsubeqs}{\begin{subequations}}
\newcommand{\esubeqs}{\end{subequations}}

\newcommand{\christoffel}[3]{\Gamma^{#1}_{\hphantom{#1}#2#3}}

\newcommand{\dd}{\mathrm{d}}                    
\newcommand{\sgn}{ \mathrm{sgn}}                

\begin{document}
\title{On a Schwarzschild-type defect wormhole}%
\vspace*{1mm}

\author{Zi-Liang Wang}
\email{ziliang.wang@just.edu.cn}
\affiliation{Department of Physics, School of Science, \\
Jiangsu University of Science and Technology, Zhenjiang, 212003, China\\}
\begin{abstract}
\vspace*{1mm}\noindent

We investigate a new type of Schwarzschild wormhole, which relies on a 3-dimensional spacetime defect with degenerate metrics. This particular wormhole is a solution of the vacuum Einstein equations. We also study the generalized Schwarzschild-type defect wormhole and discuss the Null Energy Condition. In particular, we investigate the geodesics and geodesic congruences of the generalized Schwarzschild-type defect wormhole. Additionally, we explore the optical appearance of these wormholes, shedding light on their observable features.

\vspace*{-0mm}
\end{abstract}


\maketitle

\newpage

\section{Introduction}
\label{sec:Intro}

With the nontrivial topology of spacetime, the Einstein field equations allow for a class of solutions known as wormholes \cite{visser1995lorentzian}. A wormhole has a tunnel-like structure, capable of connecting two distinct universes or two widely separated regions within the same universe.

In 1973, Ellis \cite{ellis1973ether} discovered a new type of wormhole solution. This solution is spherically symmetric and incorporates a massless scalar field with negative energy density within the framework of the Einstein field equations.  Similar solution was also found by Bronnikov \cite{bronnikov1973scalar}. Morris and Thorne \cite{Morris:1988cz} further demonstrated that this wormhole solution is traversable, allowing for instantaneous travel through space.

In the Ellis-Bronnikov-Morris-Thorne (EBMT) wormhole, the production of negative energy density requires the presence of exotic matter. However, due to the existence of such exotic matter, the stability of the EBMT wormhole is questionable. Shinkai and Hayward~\cite{Shinkai:2002gv} demonstrated that the EBMT wormhole is unstable against Gaussian pulses in either exotic or normal massless Klein-Gordon fields. Furthermore, it has been shown in Ref.~\cite{Gonzalez:2008wd} that more general static, spherically symmetric wormhole solutions of the Einstein field equations coupled to a massless ghost scalar field are unstable with respect to linear fluctuations. To avoid the need for exotic matter, one possible solution is to explore modified theories of gravity. These alternative theories of gravity may provide wormhole solutions without the requirement for negative energy density.

Recently, a proposal has been put forward by Klinkhamer~\cite{Klinkhamer:2022rsj} within the context of general relativity, which aims to eliminate the need for exotic matter in wormhole solutions. This new type of wormhole solution relies on a 3-dimensional ``spacetime defect" characterized by a locally vanishing metric determinant.   Similar type of spacetime defect has been previously proposed  to regularize the  big bang and black hole singularities~\cite{Klinkhamer:2019dzc, Klinkhamer:2019frj,Klinkhamer:2019gee,Klinkhamer:2013wla,Klinkhamer:2018ohw,Battista:2020lqv}. For more nonsingular solutions to Einstein equations involving a degenerate metric, see Refs.~\cite{Holdom:2023xip,Holdom:2023lqs}.

In this paper, we will first present an example of this type of wormhole, namely a Schwarzschild defect wormhole. This Schwarzschild defect wormhole is actually a solution of the vacuum Einstein field equations, hence the Null Energy Condition (NEC) is obviously satisfied. Furthermore, a generalized Schwarzschild-type wormhole will also be proposed  as a non-vacuum solution of the Einstein field equations. 

This paper is organized as follows: In Sec.~\ref{sec:New type of Schwarzschild wormhole}, we introduce the Schwarzschild defect wormhole, and analyze its relation to the standard Schwarzschild metric. In Sec.~\ref{sec:Generalized Schwarzschild-type defect wormhole}, we propose a generalized version of the Schwarzschild defect wormhole and discuss the NEC in this context. We also provide calculations of the curvature scalars of the generalized wormhole spacetime. Sec.~\ref{sec:geodesic and congruences} is devoted to the study of the geodesics of the generalized wormhole spacetime, including the calculation of null geodesics and the analysis of the photon sphere. We also investigate the congruences of radial geodesics. In Sec.~\ref{sec:obser}, we investigate the observational signature of Schwarzschild-type defect wormholes.  A brief summary is given in Sec.~\ref{Conclusions and discussion}. In Appendix~\ref{app:tetrad}, we present tensor calculations using the tetrad formalism. The proof demonstrating that the vacuum-wormhole metric provides a smooth solution of the first-order equations of general relativity is given in Appendix~\ref{app:vacuumsol}.

Throughout this paper, we work in reduced-Planckian units with $G = c = \hbar= 1$, where $G$ is Newton's
gravitational constant, $c$ the speed of light in vacuum and $\hbar$ the reduced Planck constant.

\section{New type of Schwarzschild wormhole}
\label{sec:New type of Schwarzschild wormhole}
A new type of traversable wormhole has been proposed recently by Klinkhamer~\cite{Klinkhamer:2022rsj}. Instead of exotic matter, this new type of wormhole solution relies on a 3-dimensional spacetime defect. In this section, we will present an example of this type of wormhole, namely a Schwarzschild-Klinkhamer (SK) defect wormhole.

The metric for the Schwarzschild-Klinkhamer defect wormhole is given as follows
\begin{align}\label{eq:SKW_metric}
  ds^2 \Big|^{\rm SK-worm} =&-\left(1-\frac{2M}{\sqrt{b^2+\xi^2}}\right)dt^2+\left(1-\frac{2M}{\sqrt{b^2+\xi^2}}\right)^{-1}\frac{\xi^2}{\xi^2+b^2}d\xi^2 \notag\\
  &+\left(b^2+\xi^2\right)\left(d\theta ^2+\sin^2 \theta d \phi^2\right)\,,
\end{align}
with real constants satisfying 
\begin{align}\label{eq:SKW_metric_condition}
  \sqrt{b^2} > 2 M\,.
\end{align}
The coordinates $t$ and $\xi$ in the metric~\eqref{eq:SKW_metric} range over $(-\infty,\, \infty)$, while $\theta \in [0,\,\pi)$ and $\phi \in [0,\,2\pi)$ are standard polar coordinates. The metric of the Schwarzschild defect wormhole~\eqref{eq:SKW_metric} resembles the metric of the regularized Schwarzschild black hole~\cite{Klinkhamer:2013wla}, but the spatial structures are different. Note that the condition $b>2M$ in \eqref{eq:SKW_metric_condition} is required to prevent the formation of an event horizon. For $M=0$, the metric~\eqref{eq:SKW_metric} reduces to the original vacuum defect-wormhole metric in Ref.~\cite{Klinkhamer:2022rsj}, see Eq.~(3.4) of that reference.  For $M=0$ and $b=0$, the metric~\eqref{eq:SKW_metric} reduces to the well known  wormhole EBMT wormhole \cite{ellis1973ether,bronnikov1973scalar,Morris:1988cz}, or Eills wormhole.   In general, the parameters $b$ and $M$ could be positive or negative, a negative $M$ could correspond to a negative gravitational mass, which may lead to ``antigravity'' effect~\cite{Klinkhamer:2018dta}. In this paper, we will mainly focus on the case $b>2M>0$.

It's worth noting that the metric given in Eq.~\eqref{eq:SKW_metric} has a degeneracy at $\xi=0$ since the determinant of the metric, $g\equiv \text{det} g_{\mu \nu}$, vanishes at $\xi=0$. Physically, the hypersurface at $\xi=0$ represents a ``spacetime defect" . The terminology  ``spacetime defect" is selected to highlight the analogy with defects found in crystals.  Just as a crystal might contain imperfections or crystallographic defects when a liquid is rapidly cooled, a spacetime defect can be seen as a remnant emerging when classical spacetime emerges from a ``quantum phase."  See Refs.\cite{Klinkhamer:2019dzc,Klinkhamer:2013wla} for in-depth discussions on the concept of spacetime defects. For a comprehensive analysis of the mathematical aspects related to degenerate metrics, we refer to Ref.\cite{Horowitz:1990qb}. Additionally, for a broader exploration of nonsingular solutions involving degenerate metrics, see Refs.~\cite{Holdom:2023xip,Holdom:2023lqs}.

As a wormhole solution, the wormhole throat is located at the defect surface $\xi=0$. The region with $\xi >0$ may correspond to the ``upper" universe, and the region with $\xi<0$ may correspond to the ``lower" universe. For further discussion on the spatial topology of the metric~\eqref{eq:SKW_metric}, see Section III B and IV B of Ref.\cite{Klinkhamer:2022rsj}.

\begin{figure}
  \centering
  \includegraphics[scale=1.0]{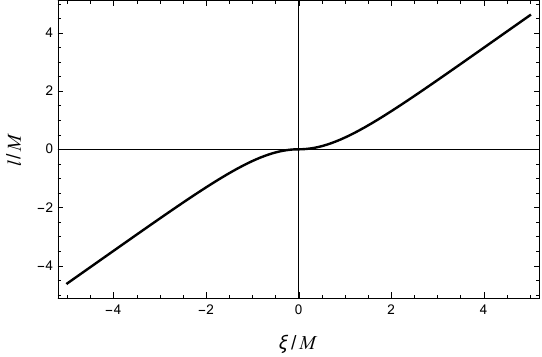}
  \caption{Proper distance from the defect throat given by Eq.~\eqref{eq:properdistance} with $b=2.5M$.}
  \label{fig:properdistance}
\end{figure}

The proper radial distance (measured by static observers) is given by
\begin{align}
  dl=\left(1-\frac{2M}{\sqrt{b^2+\xi^2}}\right)^{-1/2}\frac{\xi}{\sqrt{\xi^2+b^2}}d\xi\,,
\end{align}
from which we obtain
\begin{align}\label{eq:properdistance}
  l=&\pm\bigg[\sqrt{b^2+\xi^2-2 M \sqrt{b^2+\xi^2}}+2 M \ln \left(\frac{\sqrt{\sqrt{b^2+\xi^2}-2 M}+\sqrt[4]{b^2+\xi^2}}{\sqrt{b-2 M}+\sqrt{b}}\right)\notag\\ 
  &-\sqrt{b (b-2 M)}\bigg]\,.
\end{align}
Note that we have chosen the constant of integration in such a way that $l=0$ at the defect surface. The proper radial distance $l$ as a function of $\xi$ is plotted in Fig.~\ref{fig:properdistance}.

There is no violation of NEC in the metric \eqref{eq:SKW_metric} as it represents a solution of the vacuum Einstein field equations over a manifold with nontrivial topology. A \emph{rigorous} proof of this statement will be provided in Appendix \ref{app:vacuumsol}, where we use the first-order equations of general relativity (Palatini formalism). However, for a preliminary understanding, we can perform a quick analysis by making the following transformation:
\begin{align}\label{eq:spherical_coordiante_transfer} 
  \begin{cases}
    r=\sqrt{\xi^2+b^2}\\
    \theta =\theta\\
    \phi =\phi
 \end{cases} 
\end{align}
then the metric \eqref{eq:SKW_metric} can be written as the following standard Schwarzschild metric form:
\begin{align}\label{eq:standard Schwarz}
    ds^2 \Big| ^{\text{Schwarzschild}}   &=-\left(1-\frac{2M}{r} \right)dt^2+\left(1- \frac{2M}{r}\right)^{-1} dr^2 +r^2\left(d \theta^2+\sin^2 \theta d\phi^2 \right) \,,\\ \label{eq:standard Schwarz2}
r&\in [b,\infty) \,.
\end{align}
However, note that  the transformation Eq.~\eqref{eq:spherical_coordiante_transfer} is not a one-to-one map far from the defect, for example, $(b,\pi /2 , \pi/2)$ and $(-b,\pi /2 , \pi/2)$ in $(\xi,\theta,\phi)$ coordinates correspond to the same point, i.e.,  $(\sqrt{2}b,\pi /2 , \pi/2)$ in $(r,\theta ,\phi )$ coordinates. This observation reflects that the differential structure of the metric~\eqref{eq:SKW_metric} is different from the one of the metric \eqref{eq:standard Schwarz}. 

The Kretschmann curvature scalar for the metric~\eqref{eq:SKW_metric} is given as (the Ricci curvature scalar vanishes for vacuum solution) 
\begin{align}
 K= \frac{48 M^2}{(b^2+\xi^2)^3}\,,
\end{align}
which is well-behaved at $\xi=0$.

\section{Generalized Schwarzschild-type defect wormhole}
\label{sec:Generalized Schwarzschild-type defect wormhole}

The metric \eqref{eq:SKW_metric} can be generalized as follows:
\begin{align}\label{eq:SKW_metric2}
  ds^2 \Big|^{\rm gen-SK-worm} =&-\left(1-\frac{2M}{\sqrt{b^2+\xi^2}}\right)dt^2+\left(1-\frac{2M}{\sqrt{b^2+\xi^2}}\right)^{-1}\frac{\xi^2}{\xi^2+b^2}d\xi^2 \notag\\
  &+\left(a ^2+\xi^2\right)\left(d\theta ^2+\sin^2 \theta d \phi^2\right)\,,
\end{align}
with nonzero real constants satisfying 
\begin{align}
  \sqrt{b^2}>2M \,.
\end{align}
Again, the coordinates $t$ and $\xi$ in the metric~\eqref{eq:SKW_metric2} range over $(-\infty,\, \infty)$, while $\theta \in [0,\,\pi)$ and $\phi \in [0,\,2\pi)$ are standard polar coordinates. Still, the parameters $b$, $M$ and $a $ could be positive or negative in general, our interest will be the case with $b>2M>0$ and $a  >0$. Note that we could go back to the metric \eqref{eq:SKW_metric} if we set $a =b$.

The nonzero components of the Einstein tensor for the metric~\eqref{eq:SKW_metric2} are
\begin{subequations}\label{eq:Einstein tensor}
  \begin{align}
    E^{t}{}_{t}&=-\frac{(b^2-a ^2 ) \left(\sqrt{b^2+\xi ^2}-2 M\right)}{\sqrt{b^2+\xi ^2} \left(a ^2+\xi ^2\right)^2}\,,\\
    E^{\xi }{}_{\xi }&=\frac{(b^2-a ^2 ) \left(\sqrt{b^2+\xi ^2}-2 M\right)}{\sqrt{b^2+\xi ^2} \left(a ^2+\xi ^2\right)^2}\,,\\
    E^{\theta }{}_{\theta }&=E^{\phi }{}_{\phi }=\frac{a ^2-b^2}{\left(a ^2+\xi ^2\right)^2}+\frac{2 M \left[\frac{\left(b^2+\xi ^2\right)^2}{\left(a ^2+\xi ^2\right)^2}-1\right]}{\left(b^2+\xi ^2\right)^{3/2}}\,.
  \end{align}
\end{subequations}

Assuming that the metric~\eqref{eq:SKW_metric2} is a solution of the Einstein equations, then the energy density is given by $\rho=-T^{t}{}_{t}=-E^{t}{}_{t}/8\pi$, which is positive for $b>a$. (The energy density defined in this way is identical to that defined in the local ``proper reference frame'' \cite{Morris:1988cz}, see Appendix~\ref{app:tetrad} for more details.)  Moreover, for radial ingoing null vector (in the ``upper" universe)
\begin{align}\label{eq:null_radial_vector}
  u^{\mu}=\left[\left(1-\frac{2M}{\sqrt{b^2+\xi^2}}\right)^{-1},-\sqrt{\frac{b^2+\xi^2}{\xi^2}},0,0\right]\,,
\end{align}
we have the inequality
\begin{align}
  T_{\mu \nu} u^{\mu}u^{\nu}=\frac{b^2-a ^2}{4\pi(a ^2+\xi^2)^2}>0\; \text{for $b>a$}\,,
\end{align}
which satisfies the NEC. Note that $u^{\xi}$ in Eq.~\eqref{eq:null_radial_vector} appears to be divergent at $\xi=0$. However, this divergence disappears if we work in the local ``proper reference frame'' with tetrad formalism, see Eq.~\eqref{eq:raidal_null_vector_tetrad} in Appendix \ref{app:tetrad}. Most of the physical quantities discussed in this paper (such as $R$ and $T_{\mu \nu} u^{\mu}u^{\nu}$) are scalars, which do not depend on reference frames. Therefore, for simplicity, we work in the coordinate basis in the main text of this paper. Some non-scalar quantities calculated in the non-coordinate basis will be given in Appendix \ref{app:tetrad}.

The Ricci scalar for the metric~\eqref{eq:SKW_metric2} is 
\begin{align}
  R=\frac{4 M \left(1-\frac{\left(b^2+\xi ^2\right)^2}{\left(a ^2+\xi ^2\right)^2}\right)}{\left(b^2+\xi ^2\right)^{3/2}}+\frac{2 (b^2-a ^2 )}{\left(a ^2+\xi ^2\right)^2}\,,
\end{align}
and the Kretschmann scalar is 
\begin{align}\hspace{-8mm}
  K=&\frac{4}{\left(b^2+\xi ^2\right)^3 \left(a ^2+\xi ^2\right)^4}\cdot\Big[6 b^2 \left(b^2+\xi ^2\right)^3 \left(2 M^2-a ^2\right)+4 M^2 \left(b^2+\xi ^2\right)^2 \left(a ^2+\xi ^2\right)^2 \notag \\&+\left(b^2+\xi ^2\right)^3 \left(3 a ^4+4 M^2 \left(\xi ^2-2 a ^2\right)\right)-4 M \left(b^2+\xi ^2\right)^{5/2} \left(a ^2+\xi ^2\right)^2+3 b^4 \left(b^2+\xi ^2\right)^3\notag\\&+4 M \xi ^2 \left(b^2+\xi ^2\right)^{7/2}-12 b^2 M \left(b^2+\xi ^2\right)^{7/2}+16 a ^2 M \left(b^2+\xi ^2\right)^{7/2}+4 M^2 \left(a ^2+\xi ^2\right)^4\Big]\,.
\end{align}
Both of the curvature scalars are vanishing for $\xi \to \pm \infty$.
\begin{figure}
  \centering
  \includegraphics[scale=0.95]{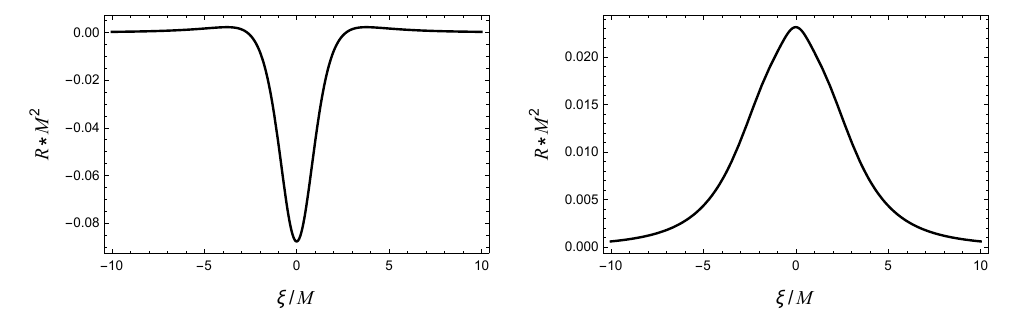}
  \caption{Ricci scalar  for $a=2M$.  At the left, we have $b=2.5M$ and $R\big|_{\xi=0}<0$. At the right, we have $b=3M$ and $R\big|_{\xi=0}>0$ }
  \label{fig:fig2}
\end{figure}

At $\xi =0$, we have 
\begin{align}
  R\Big|_{\xi=0}=\frac{4 M \left(a ^4-{b^4}\right)}{a ^4 b^{3}}+\frac{2 (b^2-a ^2 )}{a ^4}\,,
\end{align}
and 
\begin{align}
K\Big|_{\xi=0}=4 \left(\frac{4 M^2}{b^6}+\frac{3 b^2-4 b M+4 M^2}{b^2 a ^4}+\frac{3 b^2 (b-2 M)^2}{a ^8}-\frac{2 (b-2 M) (3 b-2 M)}{a ^6}\right)\,,
\end{align}
both of which are finite. Note that we have $R\big|_{\xi=0}>0$ for 
\begin{align}
  b>a \,\;\; \text{and} \; \; \frac{b^3}{b^2+a ^2} >2 M \,,
\end{align}
while $K\big|_{\xi=0}$ is always positive provided $b>a$. The Ricci scalar as a function of $\xi$ is plotted in Fig.~\ref{fig:fig2}.

With the transformation :
\begin{align}\label{eq:transformation_tilde}
 \hspace{-9mm} \tilde{\xi}=\xi\sqrt{\left[\frac{1}{\xi}\sqrt{b^2+\xi^2-2 M \sqrt{b^2+\xi^2}}+\frac{2M}{\xi} \ln \left(\frac{\sqrt{\sqrt{b^2+\xi^2}-2 M}+\sqrt[4]{b^2+\xi^2}}{\sqrt{b-2 M}+\sqrt{b}}\right)\right]^2-\frac{\tilde{\lambda}^2}{\xi^2}}\,,
\end{align}
the Schwarzschild-type defect metric~\eqref{eq:SKW_metric2} could be expressed in the   form of a general \emph{Ansatz} for the defect-wormhole proposed in Ref.~\cite{Klinkhamer:2022rsj}:
\begin{align}
  d\tilde{s}^2 \Big|^{\rm gen-SK-worm} =&-e^{2\tilde{\phi}(\tilde{\xi})}dt^2+\frac{\tilde{\xi}^2}{\tilde{\xi}^2+\tilde{\lambda}^2}d\tilde{\xi}^2 
  +\tilde{r}^2(\tilde{\xi})\left(d\theta ^2+\sin^2 \theta d \phi^2\right)\,,
\end{align}
with 
\begin{align}
  \tilde{\lambda}^2&=b^2-2Mb\,,\\
  \tilde{\phi}(\tilde{\xi})&=\frac{1}{2}\ln \left[1-\frac{2M}{\sqrt{b^2+\xi^2(\tilde
  \xi)}}\right]\,,\\
  \tilde{r}^2(\tilde{\xi})&=a ^2+\xi^2(\tilde{\xi})\,.
\end{align}
Here the function $\xi(\tilde{\xi})$ can be determined by inverting the relation \eqref{eq:transformation_tilde}. Observe that, the coordinate transformation \eqref{eq:transformation_tilde} is a $C^1$ function with a discontinuous second derivative at $\xi=0$.

\section{Geodesics and geodesic congruences}
\label{sec:geodesic and congruences}

\subsection{Geodesics and photon spheres}
In general, geodesics equations are written as 
\begin{align}
  \frac{d^2 x^{\mu}}{d\lambda^2}+\christoffel{\mu}{\sigma}{\nu}\frac{dx^{\sigma}}{d\lambda}\frac{dx^{\nu}}{d\lambda}=0
\end{align}
with $\lambda$ being the proper time for massive particle or the
affine parameter for massless particle. For our spherically symmetric defect-wormhole metric \eqref{eq:SKW_metric2},  the following  two conserved quantities could be obtained:
\begin{align}
  E&=\left(1-\frac{2M}{\sqrt{b^2+\xi^2}}\right)\frac{dt}{d\lambda}\,,\\ 
  \label{eq:define_J} J&=(a^2+\xi^2)\frac{d \phi}{d \lambda}\,.
\end{align}
Without loss of generality,  we could consider the case $\theta=\pi/2$, i.e., particles are confined to the equatorial plane. Then, the geodesic equation leads to the following constant of motion
\begin{align}\label{eq:geodesic_1}
-N\equiv \left(1-\frac{2M}{\sqrt{b^2+\xi^2}}\right)^{-1}\frac{\xi^2}{\xi^2+b^2}\left(\frac{d\xi}{d\lambda}\right)^2+\frac{J^2}{\xi^2+a^2}-{E^2}\left(1-\frac{2M}{\sqrt{b^2+\xi^2}}\right)^{-1}\,,
\end{align}
where
\begin{align}
  N &=0, \;\;  \text{for a massless particle}\,,\\
  N &=1, \;\;  \text{for a massive particle}\,.
\end{align}

\begin{figure}
  \centering
  \includegraphics[scale=0.9]{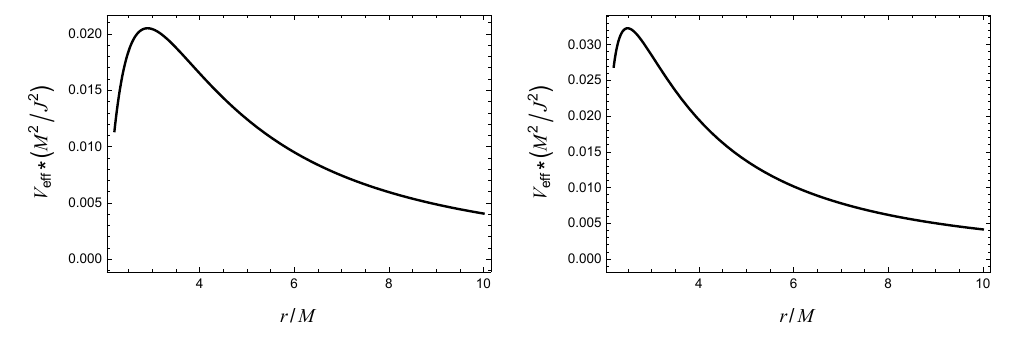}
  \caption{Effective potential for null geodesic with $b=2.2M$. At the left, we have $a=2M$ and the unstable circular orbits of photons, known as the photon sphere,  is located at $r_{\rm sh}\approx 2.9M$ ($\xi_{\rm sh}\approx 1.9M$). At the right, we have $a=1.3M$ and the photon sphere is located at $r_{\rm sh}\approx 2.5M$ ($\xi_{\rm sh}\approx 1.2M$).}
  \label{fig:Veff}
\end{figure}

By the replacement $r=\sqrt{b^2+\xi^2}$\,, Eq.~\eqref{eq:geodesic_1} can be written as 
\begin{align}\label{eq:app-radial-1d}
\frac{1}{2}  \left(\frac{d r}{d\lambda}\right)^2+ \left(\frac{1}{2}-\frac{M}{r}\right)\frac{J^2}{a^2+r^2-b^2}+\left(1-\frac{2M}{r}\right)\frac{N}{2}=\frac{E^2}{2}\,,
\end{align}
which has the same form of the equation for a unit mass particle with energy $E^2/2$ moving in a one-dimensional effective potential
\begin{align}
  V_{\rm{eff}}= \left(\frac{1}{2}-\frac{M}{r}\right)\frac{J^2}{a^2+r^2-b^2}+\left(1-\frac{2M}{r}\right)\frac{N}{2}\,.
\end{align}
For null geodesics, we have 
\begin{align}
  V_{\rm{eff-null}}= \left(\frac{1}{2}-\frac{M}{r}\right)\frac{J^2}{a^2+r^2-b^2}\,.
\end{align}
Examples of the effective potential for null geodesic is illustrated in Fig.~\ref{fig:Veff}. In what follows, we will center our attention on null geodesics and photon spheres, given their significance in observational studies. 

\begin{figure}[htbp]
  \centering
  \begin{subfigure}[b]{0.45\textwidth}
    \includegraphics[width=\textwidth]{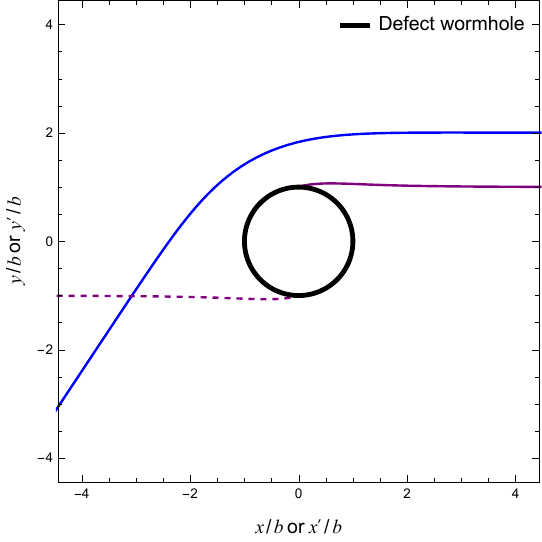}
    \caption{$b=3M$, $a=2M$.}
    \label{fig:wplot3}
  \end{subfigure} 
  \begin{subfigure}[b]{0.45\textwidth}
    \includegraphics[width=\textwidth]{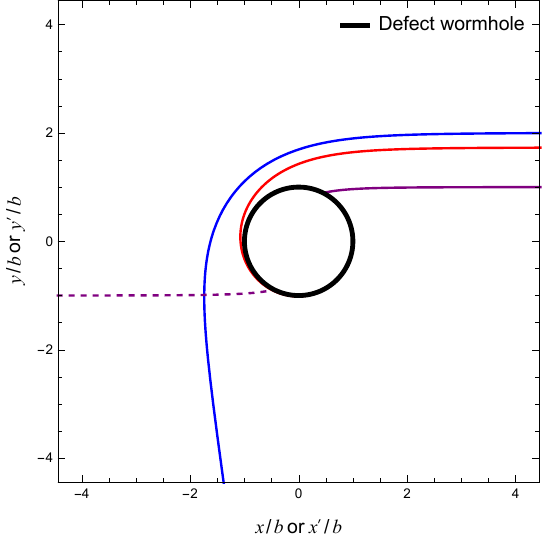}
    \caption{$b=a=3M$.}
    \label{fig:wplot2}
  \end{subfigure}
\\[1ex]
\begin{subfigure}[b]{0.45\textwidth}
  \includegraphics[width=\textwidth]{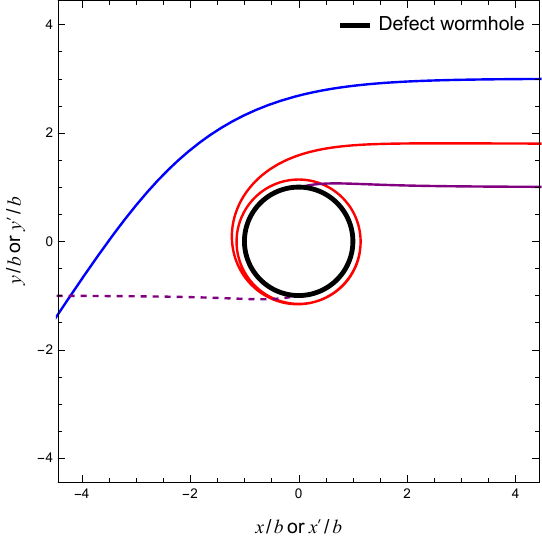}
  \caption{$b=2.2M$, $a=1.3M$.}
  \label{fig:wplot}
\end{subfigure}  
  \begin{subfigure}[b]{0.45\textwidth}
    \includegraphics[width=\textwidth]{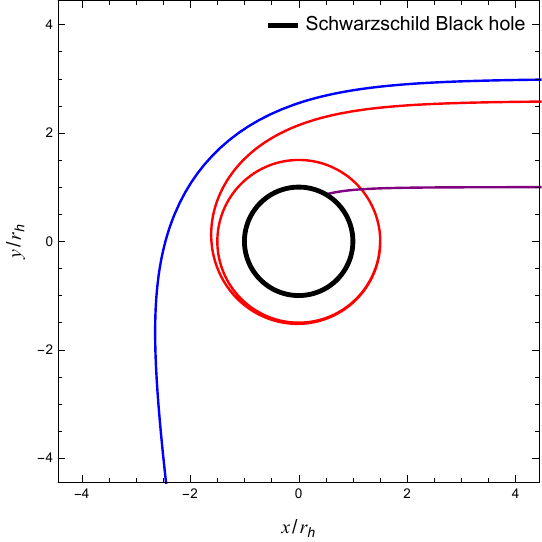}
    \caption{$b=a=0$ and $r_{\rm h}=2M$.}
    \label{fig:bhplot}
  \end{subfigure}
  \caption{(color online). Photon trajectories around (a) a defect wormhole without a photon sphere, (b) a defect wormhole with a photon sphere located at the wormhole throat $\xi_{\rm sp}=0$, (c) a defect wormhole with a photon sphere located at $\xi_{\rm sp} =1.2M$ ($r_{\rm sp}\approx 2.5M$), and (d) the Schwarzschild black hole. The Euclidean coordinates $(x, y)$ are defined on the $\theta = \pi /2$ plane by Eq.~\eqref{eq:def_xy}. In (a-c), the purple curves represent the trajectory crossing the defect wormhole, with solid line on this side of the wormhole and dashed line on the other side of the wormhole ($\xi<0$). In (d), $r_{\rm h}$ denotes the Schwarzschild radius. In (c-d), the red curves represent the trajectory of a photon approaching the photon sphere [in (b), there is strictly speaking no photon \emph{sphere}]. 
  }
  \label{fig:four_figures}
\end{figure}

Photon trajectories ($\theta =\pi/2$) around the defect wormhole is presented in~\Cref{fig:wplot,fig:wplot2,fig:wplot3}. For reference, we also present trajectories of photons around a Schwarzschild black hole in~\Cref{fig:bhplot}. In~\Cref{fig:four_figures}, we have defined the Euclidean coordinates:
\begin{subequations}
  \begin{align}\label{eq:def_xy}
  x& =r\cos \phi \,, \;\; y=r \sin \phi\,, \;\; \text{for}\;\; \xi \geq 0\,,\\
  x'& =-r\cos \phi \,, \;\; y'=-r \sin \phi\,, \;\; \text{for} \;\; \xi \leq 0\,,
\end{align}
\end{subequations}
where $r=\sqrt{b^2+\xi^2}$ and we have antipodal identification at $\xi=0$. 

The photon sphere arises from geodesic motion in the gravitational field of a compact object. It serves as a boundary between infalling geodesic bundles that scatter away to infinity and those that are captured by the object. In general, photon spheres refer to unstable circular orbits of photons that exist at the maximum of the effective potential. However, for Schwarzschild-type defect wormholes, the situation is more involved. As we will see, the existence and properties of photon spheres depend on the specific parameters in metric~\eqref{eq:SKW_metric2}. 

The extremum of the effective potential appears at $r_{\rm sh}$, where 
\begin{align}\label{eq:V_max}
  \frac{\partial V_{\rm{eff-null}}}{\partial r}\Big| _{r=r_{\rm sh}}=0\,.
\end{align}
The relevant solutions for Eq.~\eqref{eq:V_max} are summarized as follows:
\begin{itemize}
  \item $b^2-a^2>4M^2$: no solution.
  \item  $0<b^2-a^2<4M^2$ and $2M<b<3M$: \begin{align}
    r_{\rm sh}=(2\cos u+1)M\,,
  \end{align}
  where $u=\arccos[1-(b^2-a^2)/2M^2] /3\in (0,\pi/3)$\,. The existence of photon sphere also requires $r_{\rm sh}\geq b$\,, which leads to the inequality $\cos u \geq \frac{b-M}{2M}$\,.
  \item $b^2=a^2$ and  $2M<b<3M$: 
  \begin{align} r_{\rm sh}=3M\,. 
  \end{align}
  \item $a=0$ and $b=2M$: \begin{align} r_{\rm sh}=2M \,,\end{align}  and the photon sphere locates at the event horizon.
\end{itemize}
Several remarks are in order. First, for traversable wormholes, the existence of photon sphere requires at least that $0\leq b^2-a^2<4M^2$ and $2M<b\leq 3M$. In general, the photon sphere is not located at the wormhole throat, i.e., $\xi_{\rm sh}\neq 0$ (see Fig.~\ref{fig:Veff}.)

Second, it is possible for the photon sphere to be located exactly at the wormhole throat ($\xi=0$) under three distinct circumstances: 
   \begin{itemize}
    \item[i:]  $b=a =3M $\,,
    \item[ii:]  $b=0$ and $M=0$\,,
    \item[iii:]  $\cos u = {(b-M)}/{2M}$, where $u=\arccos[1-(b^2-a^2)/2M^2] /3\in (0,\pi/3)$\,.
  \end{itemize}  
  We note that in the second case, the wormhole solution reduces to the EBMT wormhole, and our analysis is consistent with the discussion presented in Ref.~\cite{Ohgami:2015nra}.

Third, although our investigation initially focused on the presence of a photon sphere, it is worth noting that the absence of a photon sphere does not pose any impediment to the existence of a defect wormhole. An illustration of such a scenario is depicted in Figure \ref{fig:wplot3}.

\subsection{Geodesic congruences}
\label{sec:Geodesic congruences}
A geodesic congruence in a subset $O$ of a spacetime manifold $(\mathcal{M}, g_{\mu \nu})$ is a family of curves such that through each point in $O$ there lies one and only one geodesic from this family \cite{wald2010general}. Geodesic congruences can be used to study the behavior of nearby worldlines in a spacetime. The evolution of a geodesic congruence can be described by the expansion $\vartheta$, the shear $\sigma_{\mu \nu}$, and the twist $\omega_{\mu \nu}$.

Consider a timelike geodesic congruence with its tangent vector field $k ^{\mu}$. Then, the expansion $\vartheta$, the shear $\sigma_{\mu \nu}$, and the twist $\omega_{\mu \nu}$ of the timelike geodesic congruence are given by \cite{wald2010general}
\bsubeqs
\begin{align}\label{eq:expansion shear and twist}
\vartheta &\equiv B^{\mu \nu} h_{\mu \nu} \,,\\
\sigma _{\mu \nu}&\equiv \frac{1}{2}(B_{\mu \nu}+B_{\nu \mu})-\frac{1}{3}
\vartheta \, h_{\mu \nu} \,,\\
\omega _{\mu \nu} &\equiv \frac{1}{2} (B_{\mu \nu}-B_{\nu \mu})\,,
\end{align}
\esubeqs
where
\bsubeqs
\begin{align}
B_{\mu \nu} &\equiv \nabla _{\nu} k _{\mu}\,,\\
h_{\mu \nu} &\equiv g_{\mu \nu} +k _{\mu} k_{\nu}\,.
\end{align}
\esubeqs
Since $B_{\mu \nu}$ is ``spatial", i.e.,
\beq
B_{\mu \nu} k^{\mu}  = B_{\mu \nu} k^{\nu} =0\,,
\eeq
we have 
\beq \label{eq:theta}
\vartheta = B^{\mu \nu} g_{\mu \nu} = \nabla _{\mu} k^{\mu}\,.
\eeq

The scalar $\vartheta$ measures the expansion of nearby geodesics in the congruence. Specifically, $\vartheta >0$ means that the geodesics are diverging, while $\vartheta <0$ means that they are converging. The quantities $\sigma _{\mu \nu}$ and $\omega_{\mu \nu}$ measure the shear and rotation, respectively, of nearby geodesics in a geodesic congruence. In this paper, we will focus on the expansion scalar $\vartheta$ as it is of particular importance in discussing the (potential) spacetime singularity. 

For a timelike radial geodesic, the four-velocity vector (with normalization condition $g_{\mu \nu} k^{\mu} k^{\nu}=-1$) is
\begin{align}\label{eq:timelike_vector}
  k^{\mu}=\left\{E\left(1-\frac{2M}{\sqrt{b^2+\xi^2}}\right)^{-1},\pm\left[\left(E^2-1+\frac{2M}{\sqrt{b^2+\xi^2}} \right)\frac{b^2+\xi^2}{\xi^2}\right]^{1/2},0,0\right\}\,.
\end{align}
Note that the divergence  at $\xi=0$ for $k^{\xi}$ in Eq.~\eqref{eq:timelike_vector} will disappear if we work in the tetrad formalism, see Eq.~\eqref{eq:raidal_timelike_vector_tetrad} in Appendix \ref{app:tetrad}.

In the ``upper" universe, the upper sign in $k^{\xi}$ of Eq.~\eqref{eq:timelike_vector}
applies to the outgoing radial geodesic and the lower sign applies to the ingoing radial
geodesic. Conversely,  in the ``lower" universe, the upper sign applies to the ingoing radial geodesic and the lower sign applies to the outgoing radial
geodesic.  Without loss of generality, we can focus on the lower sign case, which describes the scenario in which geodesics initially originate from the upper universe as ingoing geodesics and then cross the defect wormhole to propagate in the lower universe as outgoing geodesics. The expansion for such a  geodesic congruence is calculated as 
\begin{subequations}\label{eq:theta_timelike}
\begin{align}
    \theta (\xi) &=\frac{1}{\sqrt{-g}}\, \partial _{\mu} (\sqrt{-g} \xi^{\mu})\notag\\
    &= \theta _{1} (\xi)+\theta _{2} (\xi)\,,
\end{align} 
where 
\begin{align}\label{eq:theta_timelike_1}
    \theta _{1} (\xi)&=-\sgn (\xi)\frac{2 \left(E^2-1\right) }{\sqrt{b^2+\xi^2} \sqrt{\frac{2 M}{\sqrt{b^2+\xi^2}}+E^2-1}}\,, \\ \label{eq:theta_timelike_2}
    \theta _{2} (\xi)&=-\sgn (\xi)\frac{3M}{(b^2+\xi^2) \sqrt{\frac{2 M}{ \sqrt{b^2+\xi^2}}+E^2-1}}\,.
\end{align}
\end{subequations}
 The expansion is finite (for $b^2>0$) but discontinuous at $\xi =0 $. 

 Now, let us consider null geodesic congruence. We focus on null geodesics initially originate from the upper universe as ingoing radial geodesics and then cross the defect wormhole to propagate in the lower universe as outgoing radial geodesics. In this case, the normalized null vector is already given by Eq.~\eqref{eq:null_radial_vector}, and the expansion scalar of the geodesic congruence is calculated as
 \begin{align}
  \theta = -\sgn (\xi)\frac{2}{\sqrt{b^2+\xi^2}}\,.
 \end{align}
Similar to the case of timelike geodesics, the expansion of null radial geodesic congruence is also finite (for $b^2>0$) but discontinuous at $\xi =0 $.

 The observed finite discontinuities in the expansion scalar of radial geodesic congruence arise from the nontrivial topology of the manifold, and they are direct manifestations of the spacetime defect. A similar discontinuity of the expansion scalar was found in Ref.~\cite{Wang:2021rje}, where the big bang singularity of FLRW universe is replaced by a three-dimensional defect of spacetime with topology $\mathbb{R}^3$.

 \section{Observational signature}
 \label{sec:obser}
 The main purpose of this section is to investigate the optical appearance of Schwarzschild-type defect wormholes through gravitational lensing effects. We examine this phenomenon by focusing on simple scenarios where the emission originates from an optically and geometrically thin static disk situated near the wormholes. The disk is observed in a face-on orientation, and its specific intensity, denoted as $I_\nu$ (with $\nu$ representing the frequency in a static frame), is solely dependent on the radial coordinate.\footnote{While considering the face-on disk case, it is important to note, as highlighted in Ref.~\cite{Gralla:2019xty}, that more realistic scenarios involving orbiting and/or infalling matter can be considered. However, for the face-on disk configuration, these effects are degenerate with the choice of the radial profile.} 

 The specific intensity emitted from the accretion disk is denoted as $I_{\rm em}(r,\nu)$. As photons are emitted from the disk, the invariant intensity, $\mathcal{I}_{\nu} \equiv I_{\nu}/ \nu ^3$, remains conserved along their trajectories in our scenarios~\cite{Gralla:2019xty} (we neglect all possible absorptions). Therefore, the specific intensity received by the observer, $I _{\rm obs} (r_{\rm obs},\nu_{\rm obs})$, satisfies 
\begin{align}\label{eq:obs_I}
   \frac{ I _{\rm em} (r,\nu)}{I _{\rm obs} (r_{\rm obs},\nu_{\rm obs})}=\left(\frac{ \nu }{\nu_{\rm obs} }\right)^3 =\left[\frac{g(r_{\rm obs})}{g(r)}\right]^3\,,
\end{align}  
where $g(r)=(1-2M/r)^{1/2}$ is  the redshift factor. For a distant observer, Eq.~\eqref{eq:obs_I} gives 
\begin{align}
    I_{\rm obs} (r_{\rm obs},\nu_{\rm obs})=g^3 I _{\rm em} (r,\nu)\,.
\end{align}
The total observed intensity resulting from light rays emitted from a specific location $r$ is given by 
\begin{align}
    I_{\rm obs} (r_{\rm obs}) = \int I_{\rm obs} (r_{\rm obs},\nu_{\rm obs}) \dd \nu_{\rm obs}=\int  g^4 I _{\rm em} (r,\nu) \dd \nu= g^4 I(r)\,,
\end{align}
where $I_{\rm em}(r) \equiv \int   I _{\rm em} (r,\nu) \dd \nu $ is the integrated intensity at radial $r$. 

Note that when tracing a light ray backward from the observer, there is a possibility of intersecting the accretion disk, resulting in an increase in brightness due to the disk emission.  The number of times the ray intersects the disk determines the amount of brightness it accumulates. The total observed intensity is obtained by summing the intensities contributed by each intersection, as given in Ref.~\cite{Gralla:2019xty}, 
\begin{align}
    I_{\rm obs} (b)=\sum_{m}\left[g^4 I\right]\Big|_{r=r_m(b)}\,.
\end{align}
Here, $r_m(b)$ ($m=1,2,3,...$) refers to the transfer function, which denotes the radial coordinate of the $m$-th intersection position with the disk plane outside the wormholes. In numerical calculations, we consider only the first three intersections.

\begin{figure}[htbp]
  \centering
  \begin{subfigure}[b]{0.3\textwidth}
    \includegraphics[width=\textwidth]{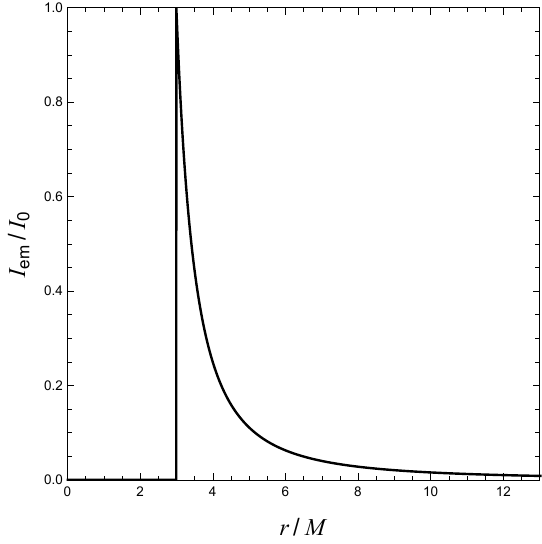}
  \end{subfigure} 
  \begin{subfigure}[b]{0.3\textwidth}
    \includegraphics[width=\textwidth]{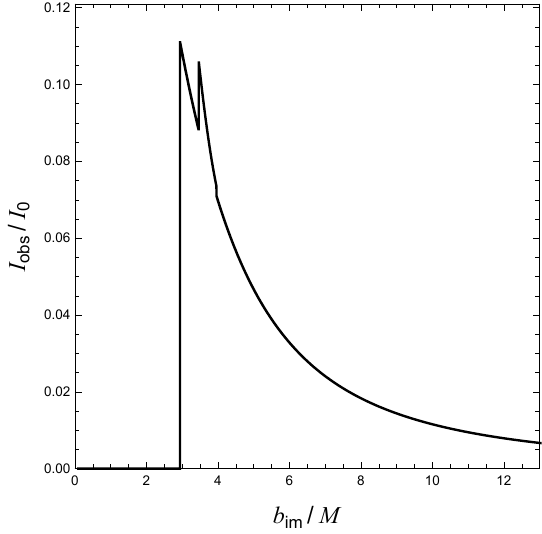}
  \end{subfigure}
  \begin{subfigure}[b]{0.37\textwidth}
    \includegraphics[width=\textwidth]{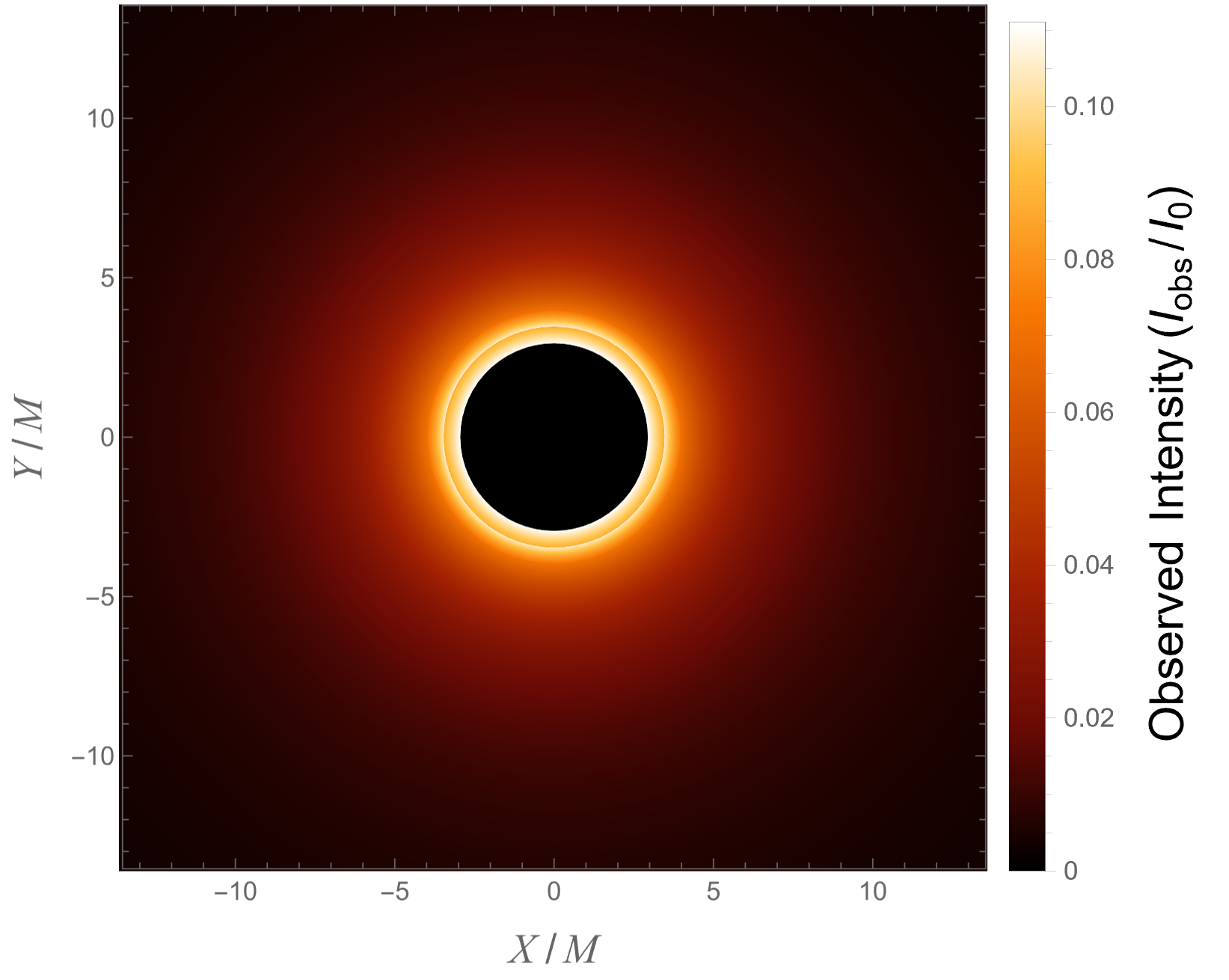}
  \end{subfigure} 
  \begin{subfigure}[b]{0.3\textwidth}
    \includegraphics[width=\textwidth]{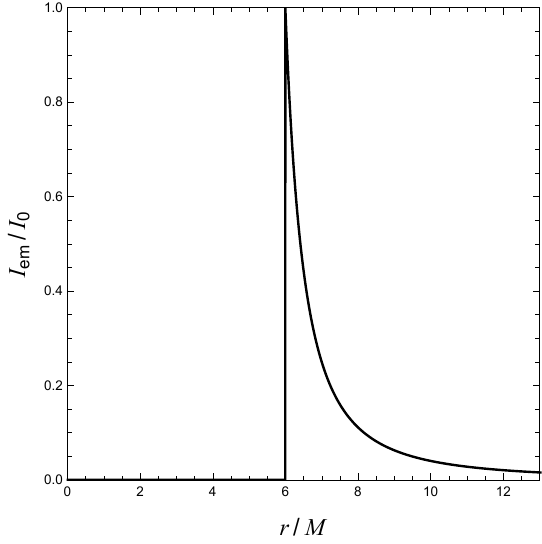}
  \end{subfigure} 
  \begin{subfigure}[b]{0.3\textwidth}
    \includegraphics[width=\textwidth]{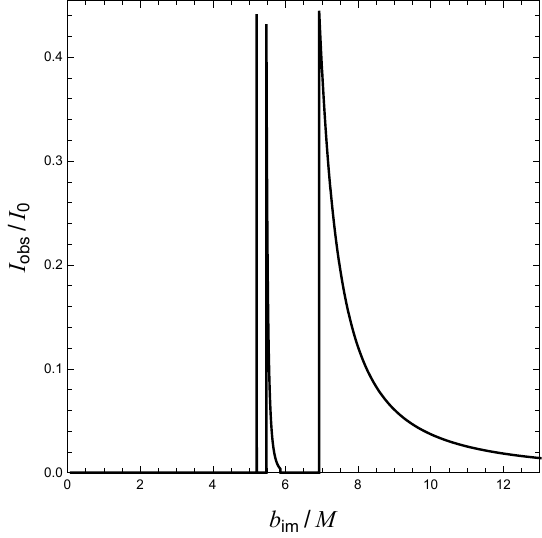}
  \end{subfigure}
  \begin{subfigure}[b]{0.37\textwidth}
    \includegraphics[width=\textwidth]{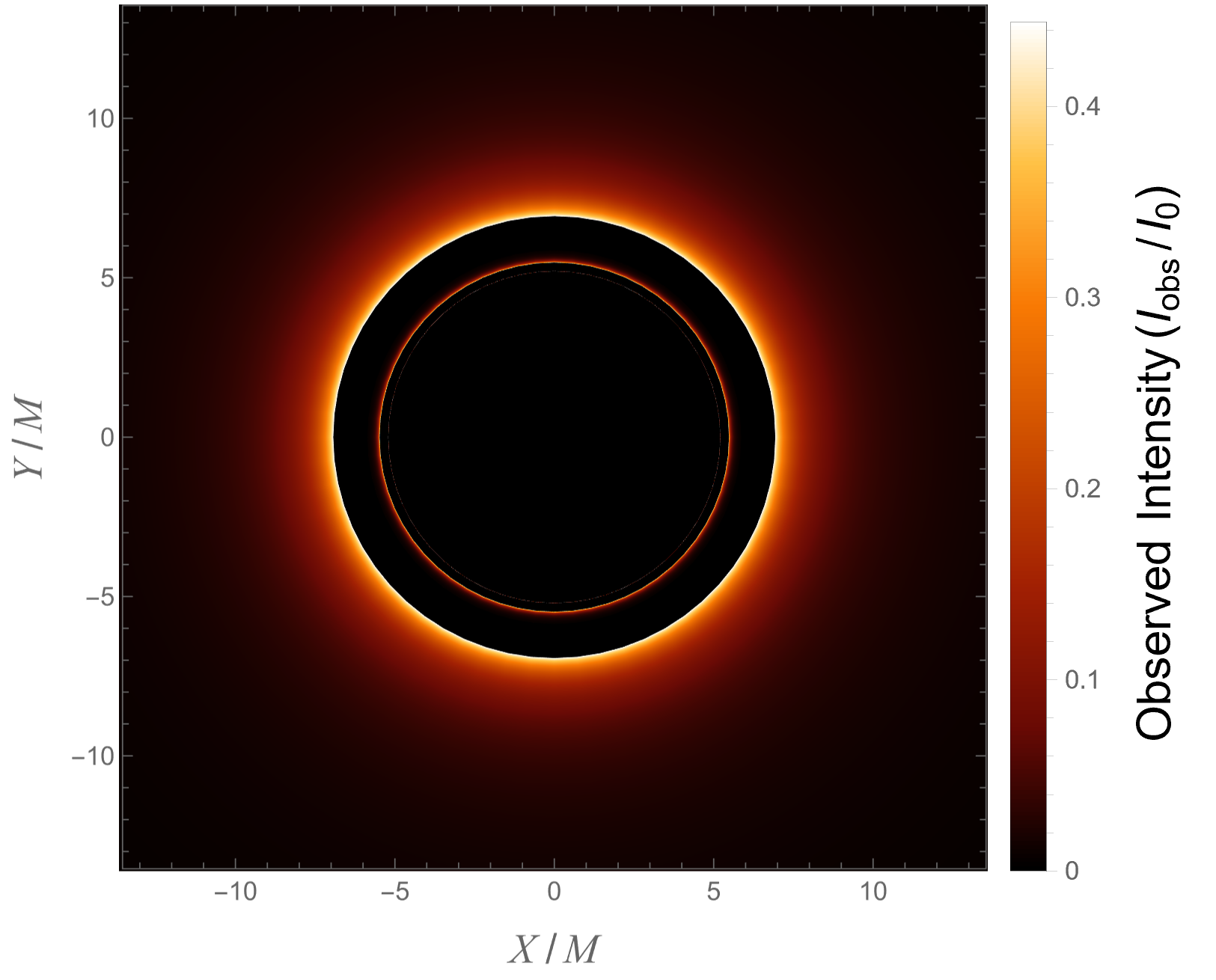}
  \end{subfigure}
  \begin{subfigure}[b]{0.3\textwidth}
    \includegraphics[width=\textwidth]{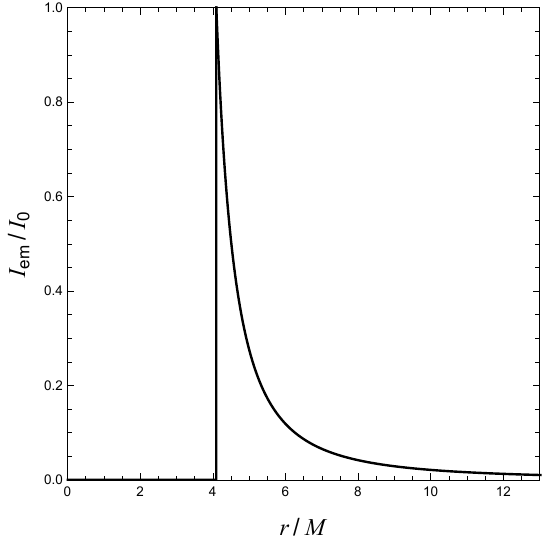}
  \end{subfigure} 
  \begin{subfigure}[b]{0.3\textwidth}
    \includegraphics[width=\textwidth]{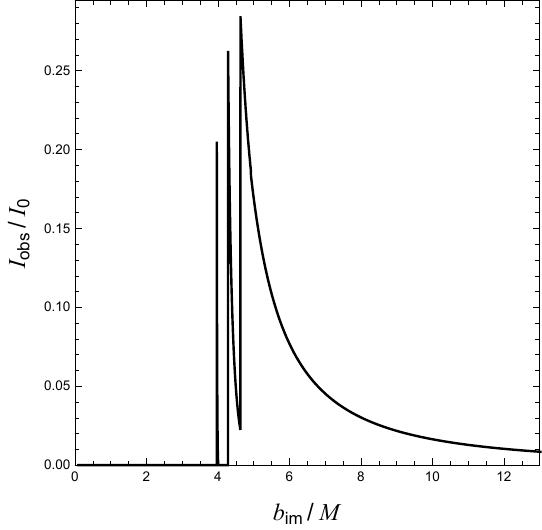}
  \end{subfigure}
  \begin{subfigure}[b]{0.37\textwidth}
    \includegraphics[width=\textwidth]{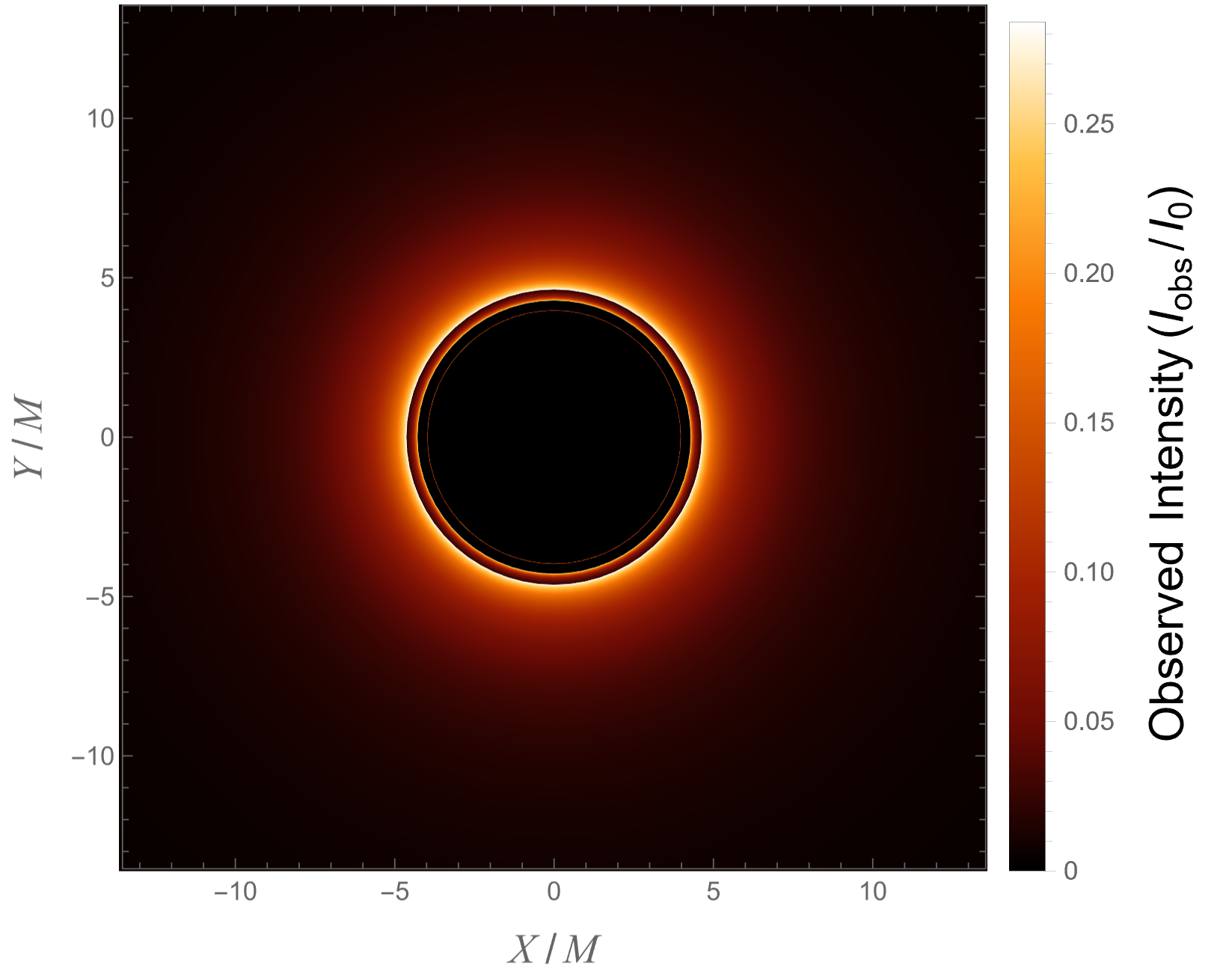}
  \end{subfigure}
  \begin{subfigure}[b]{0.3\textwidth}
    \includegraphics[width=\textwidth]{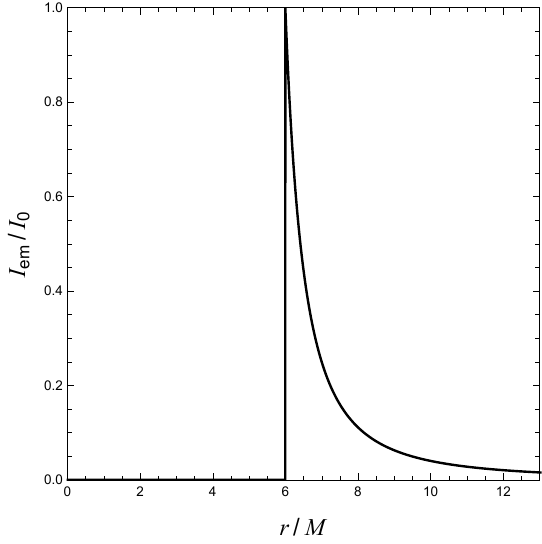}
  \end{subfigure} 
  \begin{subfigure}[b]{0.3\textwidth}
    \includegraphics[width=\textwidth]{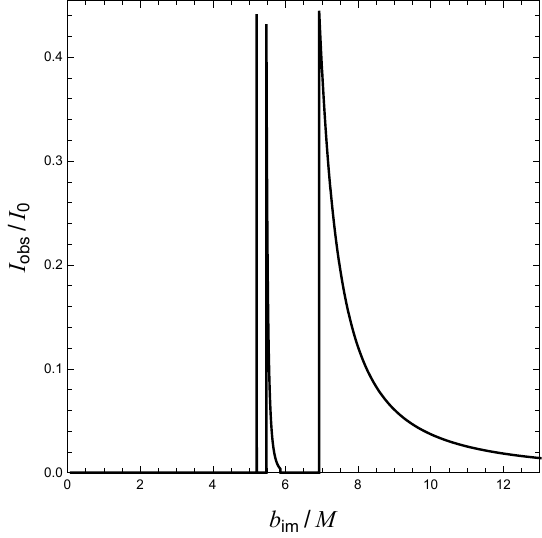}
  \end{subfigure}
  \begin{subfigure}[b]{0.37\textwidth}
    \includegraphics[width=\textwidth]{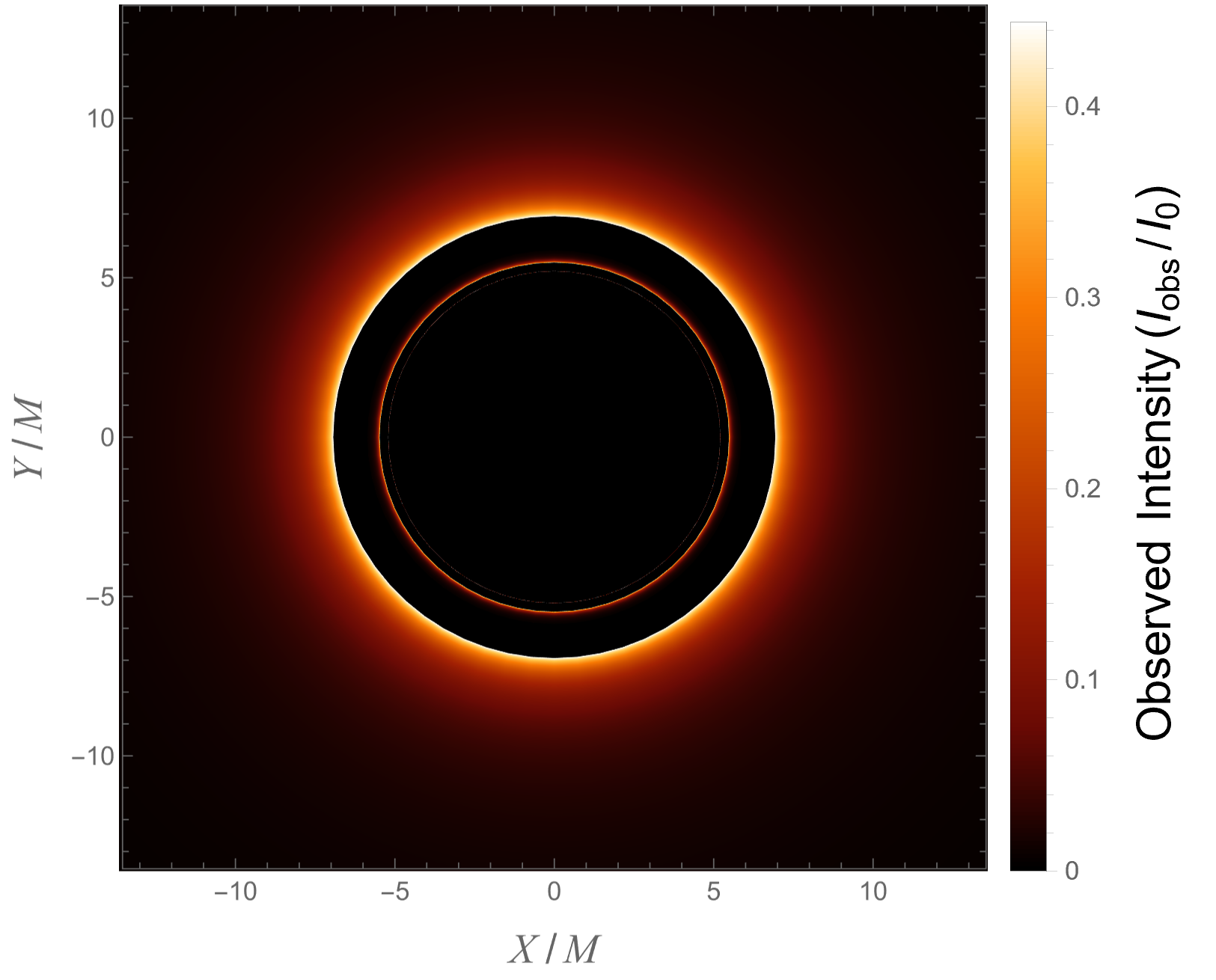}
  \end{subfigure}
  \caption{Observational appearance of an optically thin disk of emission near different spacetime geometries: (top row) a defect wormhole with $b=3M$ and $a=2M$, (second row) a defect wormhole with $b=3M$ and $a=3M$, (third row) a defect wormhole with $b=2.2M$ and $a=1.3M$, and (bottom row) the Schwarzschild black hole. The Euclidean coordinate system $(X, Y)$ describes the observational plane, which is obtained from an observer located at $r=10^5 M$.}
  \label{fig:four_figures2}
\end{figure}

For the emission intensity profile, we assume that the emission is sharply peaked near the location of the innermost stable circular orbit (ISCO),\footnote{The location of the innermost stable circular orbit is determined by the position of the local minimum in the effective potential for massive particles.} and it is modeled as follows:
\begin{align}
    I_{\rm{em}} (r) & = \left\{\begin{array}{ll}
    I_{0}\left(\frac{1}{r-\left(r_{\rm{ISCO}}-1\right)}\right)^{2}, & r\geq r_{\rm{ISCO}} \\
    0, & r < r_{\rm{ISCO}}
    \end{array}\right. \,,
\end{align}
where $I_{0}$ denotes the maximum value of the emitted intensity, and where $r_{\rm{ISCO}}$ represents the radius of the innermost stable circular orbit. Illuminated by this simplified model of the thin accretion disk, Fig.~\ref{fig:four_figures2} illustrates the optical appearance of the region near the defect wormholes. The emission intensities are presented in the left column of Fig.~\ref{fig:four_figures2}. For the defect wormhole with $b=3M$ and $a=2M$, the innermost stable circular orbit can extend up to the wormhole throat, which corresponds to $r_{\rm{ISCO}}=3M$. In the case of wormholes with $b=a$, the innermost stable circular orbit is located at $r_{\rm{ISCO}}=6M$, identical to that of a Schwarzschild black hole. For the defect wormhole with $b=2.2M$ and $a=1.3M$, we find $r_{\rm{ISCO}}\approx 4.1 M$. 

In the middle column of Fig.~\ref{fig:four_figures2}, we present the total observed intensities, which are functions of the impact parameter $b_{\rm im} \equiv J/E$ (strictly speaking,  $b_{\rm im}$ is identified as the impact parameter for null geodesics that reach infinity.) In general, the variations in observed intensities  arise from two factors: the location of the photon sphere and the location of the innermost stable circular orbit.  For instance, the observed intensity of the defect wormhole with $b=3M$ and $a=2M$ (first row) exhibits distinct characteristics compared to the other two wormholes and the Schwarzschild black hole. This disparity arises due to the absence of a photon sphere and the innermost stable circular orbit being situated at the wormhole throat. The observed intensity of the defect wormhole with $b=a=3M$ (second row) is identical to that of a Schwarzschild black hole (fourth row), as they have the same location of the photon sphere $r_{\rm sp}=3M$ and the innermost stable circular orbit $r_{\rm ISCO}=6M$. 

The differences between the optical appearances of the wormhole with $b=a=3M$ (second row) and the wormhole with $b=2.2M$ and $a=1.3M$ (third row) can be characterized by the location and height of peaks in the observed intensities. For instance, in the case of the wormhole with $b=a=3M$, the first peak (from left to right) in the observed intensity is positioned at $b_{\rm im}\approx 5.2M$, whereas for the wormhole with $b=2.2M$ and $a=1.3M$, it is located at $b_{\rm im} \approx 3.9M$.  The distinct locations of the first peaks arise from the different apparent positions of the photon spheres. Additionally, the height of the peak is primarily determined by the location of the innermost stable circular orbit. As the emission becomes more concentrated with a smaller radius, the redshift factor $g$ decreases, resulting in a lower observed intensity.

\section{Conclusion}
\label{Conclusions and discussion}

In this paper, we introduced a Schwarzschild-type defect wormhole based on recent work in Ref.~\cite{Klinkhamer:2022rsj}. The proposed defect wormhole  can be either vacuum or non-vacuum solution of the Einstein equations for parameter $a$ equal to $b$ or not. Our wormhole solution reduces to the well-known EBMT wormhole when $b=0$ and $M=0$. Notably, for $b \geq a$, the Null Energy Condition is satisfied.

Different from the case of the EBMT wormhole, in which the unstable circular orbits (known as the photon sphere) always exist and are located exactly at the wormhole throat \cite{Ohgami:2015nra},  the Schwarzschild-type defect wormhole possesses a photon sphere only if $ b^2-a ^2<4M^2$ and $2M<b<3M$. Moreover, the existed  photon sphere is  typically not located at  its throat. The exact location of the photon sphere depends on the parameters $b$, $a$ and $M$. 
Although our analysis has been primarily focused on an idealized scenario of thin disk emission, in most cases, it provides a reasonable depiction of the main differences between the optical images of a defect wormhole and that of a Schwarzschild black hole.  The location of the photon sphere plays a crucial role in distinguishing their respective images. While it is true that the observed intensity of the defect wormhole with $b = a = 3M$ is identical to that of a Schwarzschild black hole, it is crucial to note that our analysis has neglected emission originating from the other side of the wormhole. In principle, such emission (perhaps with different intensity profile) would contribute to the overall optical appearance and differentiate it from that of a Schwarzschild black hole. These aspects are left for future studies.

Returning to the metric of the wormhole as given by Eq.\eqref{eq:SKW_metric}, we observe its resemblance to the metric of the spacetime defect discussed in Ref.\cite{Wang:2022ews}. However, it is important to note that the range of $b$ and the global spatial structures of the two metrics are distinct. Their difference and connection can be reflected in the following two observations: First, the radial geodesic is complete for the defect wormhole, while not for the metric in Ref.~\cite{Wang:2022ews}. Second, the expression for the expansion scalar, given by Eq.~\eqref{eq:theta_timelike}, closely resembles Eq.~(19) in Ref.~\cite{Wang:2022ews}, with the exception of a $\sgn$ function that is present in the former.


The existence of the spacetime defect with degenerate metrics is the key assumption for the defect wormhole. Therefore, the main intriguing task for our type of wormhole is  the physical origin of this spacetime defect, which is also a crucial step to understand the nontrivial evolution of the geodesic congruences found in Sec.~\ref{sec:Geodesic congruences}.  Spacetime defects could potentially be associated with the underlying theory of ``quantum spacetime."  In loop quantum gravity~\cite{rovelli2004quantum,rovelli2015covariant}, the discrete spectra for area and volume imply the existence of a discrete ``quantum of space" and a corresponding discrete cosmological evolution referred to as a ``quantum of cosmic time."  Notably, the nonperturbative formulation of the IIB matrix model in superstring theory \cite{Ishibashi:1996xs,Aoki:1998bq} presents the intriguing possibility of emergent spacetime, potentially exhibiting spacetime defects~\cite{Klinkhamer:2020xoi}.

\section*{ACKNOWLEDGEMENTS}

It is a pleasure to thank F.R. Klinkhamer for providing valuable comments on the manuscript. This work is supported by the Natural Science Research Project of Colleges and Universities in JiangSu Province (21KJB140001) and Natural Science Foundation of Jiangsu Province (BK20220642).

\begin{appendix}
\section{Orthonormal basis (Tetrads)}
\label{app:tetrad}
The coordinate basis is chosen for tensor calculation in the main text of this paper. However, for many purposes, it turns out to be useful to work in an orthonormal basis, which is generally non-coordinate.

The orthonormal (dual) basis can be expressed in terms of the coordinate (dual) basis:
\begin{align}
  \hat{e}_{m}=e_m{} ^{\mu}\, (\partial /\partial x^{\mu}) \;\,,\; \hat{e}^m=e^m{} _{\mu} dx^{\mu}\,.
\end{align} 
The components $e^{m}{}_{\mu}$ forms a $4\times 4$ matrix (with the inverse matrix denoted by $e_m{} ^{\mu}$) satisfying 
\begin{align}
  g_{\mu \nu} =e^{m}{}_{\mu}e^{n}{}_{\nu}\,\eta_{mn}\,.
\end{align}
The $e^{m}{}_{\mu}$ are known as tetrads, or vielbeins.

For the generalized Schwarzschild-type defect wormhole Eq.~\eqref{eq:SKW_metric2}, the tetrads could be chosen as follows
\begin{equation}\label{eq:appa_tetrads}
      e^{m}{}_{\mu}=\begin{bmatrix}
       {(1-\frac{2M}{\sqrt{b^2+\xi^2}})^{1/2}}& & & \\
        &\frac{\xi}{\sqrt{\xi^2+b^2}}\left({1-\frac{2M}{\sqrt{b^2+\xi^2}}}\right)^{-1/2}& & \\
       & &\sqrt{a^2+\xi^2}& \\
       & & &\sqrt{a^2+\xi^2}\sin \theta
      \end{bmatrix}
\end{equation}
with the orthonormal basis denoted by
$
  \hat{e}_m={\partial }/{\partial \hat{x}^m}\,.
$

Working in the orthonormal basis, the radial ingoing null vector and the radial ingoing timelike vector (in the “upper” universe, i.e., $\xi >0$) are  given by
\begin{subequations}
\begin{align}
  \label{eq:raidal_null_vector_tetrad}
  u^{\hat{\mu}}=\left[\left(1-\frac{2M}{\sqrt{b^2+\xi^2}}\right)^{-1/2},-\sgn(\xi)\left(1-\frac{2M}{\sqrt{b^2+\xi^2}}\right)^{-1/2},0,0\right]\,,
\end{align}
\begin{align}
  \label{eq:raidal_timelike_vector_tetrad}
  k^{\hat{\mu}}=\left[E\left(1-\frac{2M}{\sqrt{b^2+\xi^2}}\right)^{-1/2},-\sgn(\xi)\left(E^2\left(1-\frac{2M}{\sqrt{b^2+\xi^2}}\right)^{-1}+1\right)^{1/2},0,0\right]\,,
\end{align}
\end{subequations}
both of which are finite at $\xi =0$ for $b^2>4M^2$.

The nonzero components of the Riemann tensor in the orthonormal basis are:
\begin{subequations}
\begin{align}
  R^{\hat{t}}{}_{\hat{\xi} }{}_{\hat{t}}{}_{\hat{\xi} }=&-R^{\hat{t}}{}_{\hat{\xi} }{}_{\hat{\xi} }{}_{\hat{t}}=R^{\hat{\xi}}{}_{\hat{t} }{}_{\hat{t} }{}_{\hat{\xi}}=-R^{\hat{\xi}}{}_{\hat{t} }{}_{\hat{\xi} }{}_{\hat{t}}=\frac{2 M }{\left(b^2+\xi ^2\right)^{3/2} }\,,\\
   R^{\hat{t}}{}_{\hat{\theta} }{}_{\hat{t}}{}_{\hat{\theta} }=&-R^{\hat{t}}{}_{\hat{\theta} }{}_{\hat{\theta}}{}_{\hat{t} }=R^{\hat{\theta}}{}_{\hat{t} }{}_{\hat{t}}{}_{\hat{\theta} }=-R^{\hat{\theta}}{}_{\hat{t} }{}_{\hat{\theta}}{}_{\hat{t} }=-\frac{M}{(a^2+\xi^2)\sqrt{b^2+\xi ^2}} \,,\\
   R^{\hat{t}}{}_{\hat{\phi} }{}_{\hat{t}}{}_{\hat{\phi} }=&-R^{\hat{t}}{}_{\hat{\phi} }{}_{\hat{\phi}}{}_{\hat{t} }=R^{\hat{\phi}}{}_{\hat{t} }{}_{\hat{t}}{}_{\hat{\phi} }=-R^{\hat{\phi}}{}_{\hat{t} }{}_{\hat{\phi}}{}_{\hat{t} }=-\frac{M}{(a^2+\xi^2)\sqrt{b^2+\xi ^2}}\,,\\
  R^{\hat{\xi}}{}_{\hat{\theta} }{}_{\hat{\xi}}{}_{\hat{\theta} }=&-R^{\hat{\xi}}{}_{\hat{\theta} }{}_{\hat{\theta}}{}_{\hat{\xi} }=R^{\hat{\theta}}{}_{\hat{\xi} }{}_{\hat{\theta}}{}_{\hat{\xi} }=-R^{\hat{\theta}}{}_{\hat{\xi} }{}_{\hat{\xi}}{}_{\hat{\theta} }=\frac{{M \left(-2 b^2+a ^2-\xi ^2\right)}/{\sqrt{b^2+\xi ^2}}+b^2-a^2 }{\left(a ^2+\xi ^2\right)^2}\,,\\
  R^{\hat{\xi}}{}_{\hat{\phi} }{}_{\hat{\xi}}{}_{\hat{\phi} }=&-R^{\hat{\xi}}{}_{\hat{\phi} }{}_{\hat{\phi}}{}_{\hat{\xi} }=R^{\hat{\phi}}{}_{\hat{\xi} }{}_{\hat{\phi}}{}_{\hat{\xi} }=-R^{\hat{\phi}}{}_{\hat{\xi} }{}_{\hat{\xi}}{}_{\hat{\phi} }=\frac{{M \left(-2 b^2+a ^2-\xi ^2\right)}/{\sqrt{b^2+\xi ^2}}+b^2-a^2 }{\left(a ^2+\xi ^2\right)^2}\,,\\
  R^{\hat{\theta}}{}_{\hat{\phi} }{}_{\hat{\theta}}{}_{\hat{\phi} }=&-R^{\hat{\theta}}{}_{\hat{\phi} }{}_{\hat{\phi}}{}_{\hat{\theta} }=R^{\hat{\phi}}{}_{\hat{\theta} }{}_{\hat{\phi}}{}_{\hat{\theta} }=-R^{\hat{\phi}}{}_{\hat{\theta} }{}_{\hat{\theta}}{}_{\hat{\phi} }=\frac{2 M \sqrt{b^2+\xi ^2}-b^2+a ^2}{(a ^2+\xi ^2)^2}\,,
\end{align}
\end{subequations}
All of them are finite at $\xi=0$ for $b^2>0$ and $a ^2>0$.
The Einstein tensor calculated in the orthonormal basis is identical to that given in Eq.~\eqref{eq:Einstein tensor}, e.g. $E^{\hat{t}}{}_{\hat{t}}=E^{t}{}_{t}$\,.

\section{Vacuum solution}
\label{app:vacuumsol}

The first-order (Palatini) formalism of general relativity has been recognized as highly suitable for dealing with degenerate metrics~\cite{Horowitz:1990qb,Klinkhamer:2022rsj}. In this appendix, we will demonstrate that the degenerate vacuum-wormhole metric given by Eq.~\eqref{eq:SKW_metric} represents a smooth solution of the first-order equations of general relativity.

By taking $a=b$ in Eq.~\eqref{eq:appa_tetrads}, we obtain the dual basis for the metric~\eqref{eq:SKW_metric}
\begin{subequations}
  \label{eq:app2_eq1}
\begin{align}
  \hat{e}^0&=(1-\frac{2M}{\sqrt{b^2+\xi^2}})^{1/2} dt\,,\\
  \hat{e}^1&=\frac{\xi}{\sqrt{\xi^2+b^2}}\left({1-\frac{2M}{\sqrt{b^2+\xi^2}}}\right)^{-1/2} d\xi\,,\\
  \hat{e}^2&=\sqrt{b^2+\xi^2} d\theta \,,\\
  \hat{e}^3&=\sqrt{b^2+\xi^2}\sin \theta d\phi \,.
\end{align}   
\end{subequations}
The spin-connection one-form $\omega^{m}{}_{n}$ satisfies the torsion-free condition~\cite{nakahara2003geometry}
\begin{align}\label{eq:appB_torsion-free}
  d  \hat{e}^{m} +\omega ^{m}{} _{n} \wedge \hat{e}^{n}&=0\,.
\end{align}
From the metricity condition $\omega_{m n}=-\omega_{n m}$ ($\omega_{m n} \equiv\eta _{m l} \omega ^{l}{} _{n}$), we obtain from Eq.~\eqref{eq:appB_torsion-free} the following non-vanishing components of the spin-connection one-form
\begin{subequations}
  \label{eq:app2_eq2}
  \begin{align}
  \omega^{0}{}_{1}&=\omega^{1}{}_{0}=\frac{M}{b^2+\xi ^2} dt\,,\\
  \omega^{2}{}_{1}&=-\omega^{1}{}_{2}=\left(1-\frac{2M}{\sqrt{b^2+\xi^2}}\right)^{1/2} d \theta\,,\\
  \omega^{3}{}_{1}&=-\omega^{1}{}_{3}=\sin \theta \left(1-\frac{2M}{\sqrt{b^2+\xi^2}}\right)^{1/2} d\phi\,,\\
  \omega^{3}{}_{2}&=-\omega^{2}{}_{3}=\cos \theta d \phi\,.
\end{align}
\end{subequations}
Then, from the Cartan's structure equation
\begin{align}
  d \omega ^{m}{}_{n} + \omega ^{m}{}_{l} \wedge \omega ^{l}{}_{n}&=R^{m}{}_{n}\,,
\end{align}
we obtain the non-vanishing components of the curvature 2-form 
\begin{subequations}\label{eq:app2_eq3}
\begin{align}
  R^{0}{}_{1}&=R^{1}{}_{0}=\frac{2M}{(b^2+\xi^2)^{3/2}} \,\hat{e}^0 \wedge \hat{e}^1\,,\\
  R^{0}{}_{2}&=R^{2}{}_{0}=-\frac{M}{(b^2+\xi^2)^{3/2}}\, \hat{e}^0 \wedge \hat{e}^2\,,\\
  R^{0}{}_{3}&=R^{3}{}_{0}=-\frac{M}{(b^2+\xi^2)^{3/2}} \,\hat{e}^0 \wedge \hat{e}^3\,,\\
  R^{1}{}_{2}&=-R^{2}{}_{1}=-\frac{M}{(b^2+\xi^2)^{3/2}} \,\hat{e}^1 \wedge \hat{e}^2\,,\\
  R^{1}{}_{3}&=-R^{3}{}_{1}=-\frac{M}{(b^2+\xi^2)^{3/2}} \,\hat{e}^1 \wedge \hat{e}^3\,,\\
  R^{2}{}_{3}&=-R^{3}{}_{2}=\frac{2M}{(b^2+\xi^2)^{3/2}} \,\hat{e}^2 \wedge \hat{e}^3\,.
\end{align}
\end{subequations}
It is worth noting that, at $\xi=0$,  the spin-connection one-form and curvature 2-form mentioned above are well-behaved, even though the metric is degenerate. It can be shown straightforwardly that the differential forms, given by Eqs.~\eqref{eq:app2_eq2} and ~\eqref{eq:app2_eq3}, satisfy the following first-order equations of general relativity~\cite{Horowitz:1990qb}
\begin{align}
  \hat{e}^n \wedge R^{lk} \,\epsilon_{mnlk}=0\,,
\end{align}
where $\epsilon_{mnlk}$  represents the completely antisymmetric symbol. In summary, the Schwarzschild-defetc wormhole metric from Eq.~\eqref{eq:SKW_metric}, together with the tetrad from Eq.~\eqref{eq:app2_eq1} and the spin connection from Eq.~\eqref{eq:app2_eq2}, represents a complete vacuum solution of general relativity. 
\end{appendix}
\bibliography{reference}

\end{document}